\documentstyle[11pt,aaspp4]{article}

\begin{document}
\title{The Capture of Particles by Chaotic Resonances During Orbital Migration}
\author{
A.~C.\ Quillen
}
\affil{Steward Observatory, The University of Arizona, Tucson, AZ 85721;
aquillen@as.arizona.edu}

\begin{abstract}
Because a chaotic zone can reduce the long timescale
capture probabilities and cause catastrophic events such
as close encounters with a planet or star during
temporary capture, the dynamics of migrating planets
is likely to be strongly dependent on the widths of the chaotic
zones in their resonances.  
Previous theoretical work on the resonant capture 
of particles into mean-motion resonances by orbital migration
has been restricted to the study of integrable models.  
By exploring toy 2 and 4 dimensional drifting Hamiltonian models
we illustrate how the structure in phase space of 
a resonance can be used to generalize this integrable capture theory
to include the richer phenomenology of a chaotic resonance.
We show that particles are temporarily captured into the chaotic
zone of a resonance with fixed shape and width for a time that is approximately
given by the width of the chaotic zone divided by the resonance drift rate.
Particles can be permanently captured into a drifting
chaotic resonance only if they are 
captured into a growing non-stochastic region.
Therefore resonances containing wide chaotic zones
have lower permanent capture probabilities than those
lacking chaotic zones.  
We expect large deviations from the predictions of 
integrable capture theories when the chaotic zone is 
large and the migration rate is
sufficiently long that many Lyapunov times pass while particles
are temporarily captured in the resonance.

The 2:1 mean-motion resonance with Neptune 
in the Kuiper Belt contains a chaotic zone, even 
when integrated on fairly short timescales such as a million years.
Because of the chaotic zone, the capture probability is lower than 
estimated previously from drifting integrable models.
This may offer an explanation for low eccentricity Kuiper Belt objects 
between 45-47 AU
which should have been previously captured in the 2:1 resonance 
by Neptune's migration.

\end{abstract}

\section{Introduction}

Scenarios incorporating the orbital migration
of giant planets have been proposed to
explain the orbit of Pluto and the
eccentricity distribution of Kuiper Belt Objects (\cite{malhotra}),
as well as the small orbital semi-major axes of many of the newly discovered
extra-solar planets (\cite{murray}).
A theory of resonant capture exists for well
defined adiabatically varying non-chaotic integrable 
resonant systems (similar to 
pendulums, \cite{yoder}; \cite{henrard}; \cite{henrard83}).
However, it now known that due to multiple resonance overlaps
some of the mean-motion resonances in the solar system
are in fact chaotic (\cite{wisdom}; \cite{holman};
\cite{murray98}).  From their numerical integration,
\cite{tittemore} and \cite{dermott} 
found that the chaotic nature of the resonances
did influence the capture probabilities of resonances
driving the evolution of the Uranian satellites.
\cite{tittemore} showed that the Uranian satellites were temporarily 
captured into the
chaotic zone associated with the separatrix of a resonance.
Because chaotic resonances will always temporarily capture particles
into their chaotic zones, the dynamics is fundamentally different
than that of integrable resonances where no capture takes
place unless the phase space volume of the resonance grows.
Previous theoretical work on the capture process during orbital
migration has been restricted to integrable models 
(e.g., \cite{henrard}, \cite{malhotra88}, \cite{borderies}).
In this paper, by numerically investigating toy Hamiltonian models,
we explore the general problem of capture by chaotic resonances.

In our previous work we investigated numerically the affect
of an orbiting giant planet on planetesimals interior to the planet
(\cite{quillen}).    
We showed that the strong mean-motion resonances 
captured particles and caused catastrophic events 
such as ejection by the planet or an impact with the central star.
We proposed that because impacts can enrich the metallicity of a star
at a time when the star is no longer fully convective,
the migration process offers an explanation for the high 
metallicities of stars 
with planets discovered via radial velocity searches.
The theory of adiabatically varying integrable systems
(e.g., \cite{henrard}; \cite{malhotra88}; \cite{borderies}; \cite{MD}) 
predicts permanent capture probabilities for integrable
mean-motion resonances.
However from our simulation we found that particles were often 
captured only for short periods of time.  This was particularly
evident in the simulations with slow migration rates.
We can try to understand the problem with the understanding that the strong
drifting resonances in the problem contain large chaotic zones.  
As we show below, 
when there is a chaotic zone, the probability of temporary capture 
into the resonance is 100\%.  
Because a chaotic zone is incapable of permanently capturing particles
in a drifting resonance, we expect that the probability of permanent
capture drops as the width of the chaotic zone increases.  
We expect that the distribution of temporarily capture
times and the probability of permanent capture 
depends on the width and variation
of the chaotic zone, the drift rate of the planet, 
and the distribution of diffusion timescales in the zone.


\section{The Forced Pendulum Analogy}

These phenomena can be illustrated simply with a mathematical model
for the forced pendulum.
This is a simple two dimensional Hamiltonian system which exhibits 
the chaos caused by resonant overlap (e.g., \cite{holman}).
The Hamiltonian is
\begin{eqnarray}
H(P,\psi) 
     &=& {1 \over 2} P^2 + K\left[1 + a \cos(\nu t)\right]\cos(\psi) \\
     &=& {1 \over 2} P^2 + K\cos(\psi) + {Ka \over 2} \left[
 \cos(\psi+\nu t)+ \cos(\psi-\nu t) 
 \right] 
\nonumber
\end{eqnarray}
and has resonances at $p = 0,\pm \nu$ with $\psi = \pi$.
To drift the resonance we modify the Hamiltonian
\begin{equation}
H (P,\psi) = {1 \over 2} P^2 + K\left[1 + a \cos(\nu t)\right]\cos(\psi)  + b P.
\label{Ham_2D}
\end{equation}
For the resonance to migrate or drift, 
$b$ must be a function of t.  
We can allow the resonance to grow without changing its shape if
we set $\dot{a}=0$ and $\dot{K}>0$..

In Figure 1 we show the result of drifting two different systems, one
with a large chaotic zone and the other without.
We numerically integrated the above Hamiltonian 
using a conventional Burlisch-Stoer numerical scheme.
Both numerical integrations pertain to systems with constant resonant widths 
and shapes ($\dot{K},\dot{a} = 0$).
Since the size and shape of the resonance does not change with time,
the capture probability predicted for a non-chaotic resonance is zero
(e.g., \cite{henrard}).  However when the chaotic zone is large,
particles can spend a significant amount of time in the chaotic
zone associated with the separatrix before passing to the other side
of the resonance.  
%
We can define an effective width for the chaotic zone as
$\Delta P_z  = {V_z \over 2\pi}$ where $V_z$ is the volume
in phase space of the chaotic zone connected to the separatrix.
Particles spend different time periods temporarily captured
in the chaotic zone.
The distribution of temporary capture times has a lower
edge at nearly zero time in the resonance. 
The mean length of time captured into the resonance is 
\begin{equation}
\Delta T_c = \Delta P_z / |\dot{b}| 
\end{equation}
when the shape of the resonance is held fixed.

Because the chaotic zone may contain regions with different
diffusion timescales, we expect that the final particle
distribution will have a dependence on drift rate.
To explore this we performed integrations for 3 different
drift rates for the system shown on the right hand side
of Figure 1.
The final particle probability distributions are shown in Figure 2.
For the faster drift rates, the final particle
distribution is nearly flat, however for the slower
drift rates, the distribution is more triangular and has
a longer tail.  In the longer integrations we expect that particles
can be trapped in regions with longer diffusion timescales.

\subsection{The Probability of Capture}

As shown by \cite{yoder} and \cite{henrard}, we can estimate 
the capture probability for an adiabatically drifting resonance
by considering the volume of phase space per unit time
which is passed by the separatrices of the resonance.
We are referring to 
a system with 2 separatrices (the non-chaotic example
shown in Figure 1).
For $P_+(\psi,t)$ and $P_-(\psi,t)$ the momenta of the 
upper and lower separatrices as a function of angle and time,
the rate of volume swept by the upper separatrix is $B_+$  
where 
$B_+ \equiv \int_0^{2 \pi} {d \over dt} P_+(\psi,t) d\psi$.
The growth rate of phase space volume in the resonance
is $B_+ - B_-$, where 
$B_-$ is the corresponding expression for the lower separatrix.
Particles swept up by the resonance must either be captured or ejected.
The capture rate depends on the ratio of volume increase
in the resonance to that swept up by the resonance and so is 
given by 
\begin{equation}
P_c = (B_+ - B_-)/ B_+.
\end{equation}

This is shown with more rigor by 
\cite{henrard} for the drifting pendulum 
(Hamiltonian in equation (\ref{Ham_2D}) restricted to $a=0$).
The probability of capture
\begin{equation}
P_c = 
\left\{
\begin{array}{ccl}
f && {{\rm if} ~ 0<f<1;}  \\
1 && {\rm if} ~ f \geq 1 
\end{array}
\right\}
\end{equation}
where 
\begin{equation}
f = {2 \over 1 - {\pi \over 2}{\dot{b} \over \dot{K}}\sqrt{K}}.
\label{fequal}
\end{equation}
We have assumed that particles start at large $P$ and
the growing resonance ($\dot{K} > 0$) drifts upwards ($\dot{b} <0$).
If the resonance shrinks, $\dot{K} \leq 0$, then the permanent capture
probability is zero.
We have computed permanent capture probabilities 
for a series of systems with different sized chaotic zones
by choosing different values for $a$ in each system. 
Since we do not allow $a$ to vary in each individual simulation, 
the shape (not size) of the resonance remains the same while
the resonance drifts.
For these simulations we set $\dot{K}$ such that when $a=0$
the probability of capture is $P_c = 1$.  
We see in Figure 3 that
the permanent capture probability of the resonance 
drops when it has an decreasing volume fraction covered
by integrable motion or stable islands.
The fraction of the resonance covered by islands drops exponentially
as $a\to 0$ so the capture probability exponentially 
approaches 1 for small $a$.

When the resonance contains a chaotic separatrix, we expect
its permanent capture probability to differ from that of
a resonance of similar shape lacking a chaotic separatrix.  
In the limit of an entirely
chaotic resonance, the resonance cannot permanently
capture particles 
unless the resonance width grows faster than the drift rate
and the resonance is effectively stationary.
A drifting resonance requires a stable, integrable, non-chaotic, 
growing island to capture particles.
We expect the capture probability to be given by 
\begin{equation}
P_c ={\dot{V_i}\over B_+}.
\end{equation}
where $\dot{V_i}$ is the growth rate of the island or
islands of non-chaotic  phase space
volume, and $B_+$ is the rate that particles are swept into
the chaotic zone of the resonance.
When the volume of the chaotic zone shrinks to zero we recover the formalism
of the integrable model.

\subsection{Escape from the 2:1 resonance in the Kuiper Belt}
\cite{malhotra} proposed that Neptune's outwards migration
was responsible for the capture of Pluto and other Kuiper
belt objects into the 3:2 and 2:1 mean-motion resonances
with Neptune.  Because the capture
probability predicted with the integrable formalism 
is 1 for low eccentricity objects, particles are not
expected to escape the 2:1 resonance.
The existence of low eccentricity objects 
between 45-47 AU, just within the 2:1 resonance,
has posed a challenge to explain.  
Nevertheless, the numerical simulations
of the migrating major planets by \cite{malhotra} and \cite{hahn} 
showed that particles with initially low eccentricity could pass 
through this resonance.  \cite{hahn} suggested 
that the stochastic migration rate of Neptune seen
in their simulations might result
in particles leaving the 2:1 resonance, however
the simulation with a smooth migration by \cite{malhotra} still showed
that low eccentricity particles could avoid capture.
Numerical integrations have shown that on long timescales 
(4 Gyrs) the 2:1 is possibly entirely chaotic (\cite{renu2000}). 
Based on our understanding of our toy problems we may be able to offer
an alternative explanation for passage of particles through
the 2:1 resonance even when Neptune's migration is smooth. 

The numerical simulations of the 2 dimensional toy model 
we discussed above were in the regime 
$\nu \sim \omega_0$ where $\omega_0 = \sqrt{K}$ (see equation (\ref{Ham_2D})).  
This regime is appropriate for
the asteroid belt (\cite{holman}) 
and for the simulation of inwards orbital migration of 
a Jovian sized giant planet (as by \cite{quillen}).
However in the Kuiper Belt secular oscillation frequencies
are slow compared to the mean-motion resonance
oscillation frequencies. 
If the simple model described by equation(\ref{Ham_2D}) 
were appropriate we would
expect $\nu \ll \omega_0$ and that the resonances are highly overlapped.
In this regime on long timescales the resonance is entirely chaotic.
However, on short timescales 
we can consider the resonance to be a slowly
varying integrable system.
In this case we could use the formalism for calculating capture for
slowly varying separatrices (e.g., \cite{haberman}; \cite{neishtadt}).

Even though the secular oscillations frequencies are slow
in the Kuiper belt, the edges of the resonances are still
found numerically to exhibit chaotic motion even
on the short timescales of millions of years
(e.g., \cite{morbidelli}, \cite{renu2000}). 
So the theory of capture in the regime of slowly varying separatrices,
is still not applicable to the theory of capture
by Kuiper Belt resonances.
The numerical integrations (e.g. \cite{morbidelli} on the 3:2
and \cite{renu2000} on the 2:1 resonances) show that the problem is
not as simple as a 
simple model of multiple resonances restricted to a few terms
and that even on short timescales, the resonances contain
fairly wide chaotic zones.

When a resonance contains regions with widely different diffusion
timescales, the problem is more complicated than illustrated with
our simple 2D model explored above.  However if the migration
rate is fast compared to the diffusion timescale in a particular
region of phase space, we can consider that region
to be integrable.  This is equivalent to modeling the resonance
as a series of overlapped resonances with different frequencies, 
and ignoring the terms with the slowest frequencies.

If a particle can remain in a region of libration
for $10^6-10^7$ years then for orbital migration rates such
that the resonance is passed on this timescale 
($\Delta a/\dot{a} \sim  10^6-10^7$ years,
for $\Delta a$ the width of the resonance and $\dot{a}$ the resonance 
migration rate)
we can consider that region to be integrable, and so capable
of capturing particles.
For low initial particle eccentricities, $e_{init} < 0.1$, \cite{renu2000}
found that about a fifth of phase space was likely to be chaotic
in the 2:1 resonance on fairly short timescales.
In other words, \cite{renu2000} found no regions of stable libration for
about 1/5 of possible values for the resonant angle.
Even though the integrable model predicts a 100\% capture probability
for $e_{init} < 0.06$, if 1/5 of the resonance is chaotic,
then the capture probability is likely to be only $\sim 4/5$
for migration rates $\sim 10^6 - 10^7$ years.
This is one way that 
low eccentricity objects could pass through the 2:1 resonance, and
a possible explanation for low eccentricity Kuiper Belt objects
between 45-47 AU
which should have been previously captured in the 2:1 resonance
by Neptune's migration.

Following a period of fairly swift migration, it is likely that
Neptune underwent migration at a slower rate.
For slower migration rates we can consider a larger fraction
of the 2:1 resonance to be chaotic and the capture probability
will be even smaller.  Particles
pumped to high eccentricity while caught in the resonance, 
can subsequently escape.
In our previous simulations (\cite{quillen}) we saw that
particles could be temporarily captured, have their eccentricities
increase while in the resonance 
and then could escape with eccentricities that were
not neccessarily well above their initial value.
Any objects released at high eccentricity from the 2:1 resonance in the Kuiper
Belt would not be observed today.  
Because of their high eccentricities, following escape from the resonance
they would be likely to suffer encounters with Neptune.
Low eccentricity particles which escaped the 2:1 resonance 
are likely to remain in stable orbits for the age of the solar system.
The remaining objects we expect to find today should 
either reside in stable high eccentricity regions of the 2:1 resonance
or at low eccentricity inside the semi-major axis of this resonance.  
As the observational constraints on the orbital
elements of Kuiper Belt objects improve, we expect it will
be possible to numerically explore the validity of this kind of scenario.
Because of the different different diffusion timescales in the resonance,
such a study may provide constraints on Neptune's migration rate 
as a function of time.

\section{The Four Dimensional Analogy}

The celestial dynamics problem restricted to the plane containing
the planet and sun is a 4 not 2 dimensional problem.
Oscillations in semi-major axis are coupled to those in
eccentricity.  While a particle is temporarily captured into
a resonance, large excursions in eccentricity can also take place
(as discussed by \cite{tittemore} and \cite{dermott}).

To illustrate what happens in the 4 dimensional system
we explore a toy model with Hamiltonian similar to that given
in equation (27) of \cite{murray97}
\begin{eqnarray}
H(P,\psi;I,\phi) &=&
 {1 \over 2} P^2 + K I^q \left[1 + a \cos(\phi)\right]\cos(\psi)  + b P
+ \nu I  \\
&=&
 {1 \over 2} P^2 + K I^q \left[\cos(\psi) 
+ {a\over 2} \cos(\psi + \phi)
+ {a\over 2} \cos(\psi - \phi)
\right]  + b P
+ \nu I 
\nonumber
\end{eqnarray}
In the celestial dynamics problem,
$I$ is primarily related to the eccentricity of the particle,
$P$ to the semi-major axis,  $\psi$ corresponds to resonant
angle and $\phi$ to the longitude of perihelion.
$\nu$ corresponds to the precession frequency and is given
by secular theory, and $q$ depends on the order of the resonance.
In the celestial dynamics problem, 
$K$ and $a$ instead of being constants would be functions of 
the semi-major axis or $P$.  If we do
a cannonical transformation to variables $\Gamma =kI$; $\theta = (\psi-\phi)/k$
for $k=2q$
then we recover a Hamiltonian in the form
$H  \propto
 {1\over 2} \Gamma^2 + b' \Gamma + K' \Gamma^{k/2} \cos(k\theta) + ..$.
This Hamitonian is the integral model 
often used to estimate capture probabilities 
into k'th order mean-motion resonances (e.g., \cite{henrard}; \cite{malhotra88};
\cite{borderies}; \cite{MD}). 

An integration of the 4 dimensional Hamiltonian given above 
with initial conditions $I=1$ so that
it is close analogy with the two dimensional analog discussed above,
yields phenomenology (see Figure 4) remarkably similar to that we observed
in our orbital migration numerical integrations (\cite{quillen}).
We see in Figure 4 that when temporary capture takes places 
there tends to be an increase in the mean value of $I$.  
After capture $I_{mean} \sim (H_0-1/2P_0^2)/\nu$ where $H_0$ is the value
of the Hamiltonian and $P_0$ is the centre of the resonance
when capture takes place.  
Since excursions in $I$ can take place, resulting in a wider
resonance, particles can remain in the resonance for longer
times than possible in the two dimensional analog.  The 
resulting momentum probability distributions are broader (shown
in Figure 5), particularly when the migration rates are slow.

Because $I$ is related to the particle eccentricity,
excursions in $I$ represent excursions in
eccentricity which could result in catastrophic events
such as encounters with a planet or star.
If the eccentricity undergoes a random walk 
once a particle is captured into the resonance
(as discussed in the diffusion model by \cite{murray97}), then
when the capture time is long compared
to the diffusion timescale, catastrophic events would be more
likely to occur.  
In this situation the dynamics 
would be strongly dependent upon the planet's
orbital migration rate.
\cite{murray97} discusses the diffusion timescale in
terms of the Lyapunov time which for the simulations shown
in Figures 4 and 5 is $T_L \sim 2\pi$.
The slower drift rate integrations shown in Figure 5 
represent cases where particles are trapped for many
Lyapunov times.  This may explain why 
their final momentum probability distributions 
exhibit larger tails.

We now discuss the phenomenology seen in our previous
numerical simulations (\cite{quillen}).
Though initial particle eccentricities were fairly low, $e_{init} \sim 0.1$, 
because we chose planet eccentricities of $e_p = 0.3$ for many of
the simulations, the forced eccentricities were high and the mean
initial eccentricities tended to be $\sim 0.05-0.3$.
The integrable model (\cite{malhotra88}) predicts
a 100\% capture probability for a Jupiter mass planet into the 3:1
resonance for eccentricities less than $\epsilon < 0.07$.
Therefore we expect fairly low permanent capture probabilities
of $0.2-0.5$ into this resonance (\cite{borderies}).

For the 3:1 mean-motion resonance with
a Jovian sized planet, the Lyapunov timescale (which is related to
the secular oscillation frequency, \cite{holman}) is about $10^3$ times
the planet period.  For a resonance width that is 
about 0.02 the radius of the planet, we expect temporary
capture to last a time of $2 \times 10^4$ planet periods which is
only 20 Lyapunov times
for a migration rates $D_a \equiv P \dot{a}/a =  10^{-6}$
given in units of the initial planet orbital period. 
For most of the simulations, temporary
capture times were fairly short and deviations in eccentricity caused
by temporary capture into the chaotic zone were limited.  
We also did 2 simulations with slower 
migration rates of $D_a = 3 \times 10^{-7}$.  In these simulations
particles are expected to remain in the chaotic zone for longer,
$\sim 60$ Lyapunov times.  In the slower migration simulations
we saw more examples of resonances temporarily capturing particles. 
The probability of capture into the 3:1 (including temporary captures)
was larger than that predicted from the integrable theory 
(using rough numbers from Table 2 of \cite{quillen}).
The simulations showed
that some of the temporary captures resulted in catastrophic events
such as encounters with the planet or star.
As expected from the final momentum distributions shown in
Figure 5,
in the slower simulations, the passage of the resonances heated the
eccentricities and semi-major axis distribution
to a larger extent than in the faster migration rate simulations.

We conclude that the chaotic zone in
is likely to have the strongest affect for slow migration rates.  
We expect large deviations from the predictions of
integrable capture theories when the chaotic zone is   
large and the migration rates is 
sufficiently long that many Lyapunov times pass while particles
are temporarily captured in the resonance.

In the simulations shown in Figures 4 and 5, because 
$K$ and $a$ are constants, the size of the resonance does not
change as $P$ varies.
If we allowed $K$ and $a$ to be functions of $P$ we could
allow the size of the resonant islands to grow as the resonance
drifts.  This would then would allow the integrable islands in
the resonance to capture particles for longer periods of time.
The separation between overlapping resonances 
(here described by $\nu$) is given
by secular theory (e.g., \cite{holman}) and so will not vary 
quickly once a particle is trapped in one of the integrable islands.  
However, as the eccentricity grows, the widths of the individual
sub-resonances will also grow and the resonances will
overlap to a larger degree.  We therefore expect 
the size of the chaotic region to grow after a particle is trapped
in the resonance.  
Because the volume in the integrable islands will at some time  
shrink instead of growing, particles will eventually
escape the resonance. 
This provides a complete analogy for the process 
of resonant capture and escape that we saw in our previous simulations
(\cite{quillen}).

\section{Summary and Discussion}

By exploring toy Hamiltonian systems we have shown
how the capture process is fundamentally different
for drifting chaotic resonances than for 
drifting integrable systems.
Previous theoretical work on resonant capture has been limited to
integrable models.  In this paper, we have illustrated how
an understanding of the structure in phase space 
of a resonance can be used to generalize this integrable theory
to include the richer phenomenology of a chaotic resonance.

1) We have shown that particles are temporarily captured into the chaotic zone
of a drifting resonance.
For a resonance of fixed shape,
the capture time depends on the effective width of the zone
and the drift rate.  
In fact, temporary capture will take place even when the resonant width is
shrinking.  This is not true in the case of an integrable resonance.

2) The permanent capture probability of a resonance
depends on the ratio of phase space volume growth rate 
in its integrable islands
compared to that swept up by the resonance.  
This implies that permanent capture probabilities are lower for
resonances containing larger chaotic zones
than those estimated from drifting integrable models.
This offers a possible explanation for the passage of particles
through the 2:1 resonance in the Kuiper Belt following migration
by Neptune, and for the temporary captures
seen in our previous simulations (\cite{quillen}).

The continued migration of a planet via ejection of planetesimals
depends on the fraction of particles remaining after a strong
resonance has swept through the disk.
Since catastrophic events such as close encounters
with a planet or star can take place
during temporary resonant capture, and because particles
are more likely to escape drifting chaotic resonances, the dynamics 
of migrating systems may be strongly influenced by this process.
We expect the largest deviations from the predictions of
integrable capture theories when the chaotic zones are large
and migration rates are
sufficiently long that many Lyapunov times pass while particles
are temporarily captured in the resonance.

3)  The passage of chaotic resonances results in a heating of 
the particle momentum distribution.
The momentum distribution width is increased  
by the effective momentum width of the chaotic zone.  
The final momentum distribution shape is dependent on the drift rate.

In this work we have numerically explored some simple drifting
Hamiltonian systems.   We suspect that the final momentum distributions
of a drifting chaotic resonance is sensitive to the
the distribution of diffusion times in the resonance.
One way to explore this possibility would be to numerically
investigate toy models with different structure in their chaotic 
zones.

In future, by exploring in detail the timescales and 
structure of solar systems resonances, and comparing the results 
of numerical simulations with the observed
distribution of objects, we suspect that the form of
orbital migration of the planets may be constrained.
The study of drifting chaotic
resonances will also be applied to other fields.
Resonances play an important role in the
stellar theory of dynamical friction.  The possibility that 
additional heating can be caused by chaotic resonances has 
not yet been explored.
The brightness of some zodiacal and Kuiper Belt dust belts depends
on the lifetime for dust particles to remain in resonances.
We expect that studies which take into account the phase space structure of 
solar system resonances may be used to derive better
estimates for the lifetime of dust particles in these belts.

\acknowledgments

This work could not have been carried out without helpful
discussions with Matt Holman, Renu Malhotra, Bruce Bayly, Andy Gould, 
Mark Sykes, Elizabeth Holmes, and John Stansberry.
Support for this work was provided by NASA through grant numbers
GO-07886.01-96A and GO-07868.01-96A
from the Space Telescope Institute, which is operated by the Association
of Universities for Research in Astronomy, Incorporated, under NASA
contract NAS5-26555.

\clearpage

{}

\vfill\eject

\begin{figure*}
\vspace{10.0cm}
\caption[junk]{
Forced pendulum numerical integrations.  
Integrations have $K=0.17$ and $\nu=1$ (see Hamiltonian
given in equation 2).
The horizontal axes are the angle $\psi$ and the vertical axes are 
the momentum $P$.
Points are plotted every $t= 0$ modulo $2\pi/\nu$.
a) The structure of the fixed stationary resonance ($\dot{b}=0$) with $a=0.001$.
b) The structure of the fixed stationary resonance with $a=0.2$.
c) The resonance with $a=0.001$ 
was begun centered at the location shown in the upper
figures and then allowed to drift upwards at
at a rate of $dP/dt = -\dot{b} \sim 10^{-3}$.
20 particles were integrated with initial 
conditions $P \approx 2.0, \psi \approx \pi$.
The final orbits of the 20 particles are shown.
d) Same as c) but for the system with $a=0.2$.
Since the system on the left (shown in a) and c)) is not chaotic, the final
particle distribution is identical to that of the initial
one only shifted by the effective width of the resonance 
(the volume of the resonance divided by $2 \pi$).
Since the resonance width is held
fixed, particles are not captured into the resonance but
simply jump from one side of the resonance to the other.
Because the system on the right (shown in b) and d) contains
a large chaotic zone,  
particles are temporarily captured into
the chaotic zone.   The final particle momentum distribution has
been heated and has width approximately equal
to effective width of the chaotic zone.
The length of time captured is directly related to the final
momentum reached.
}
\end{figure*}

\begin{figure*}
\vspace{12.0cm}
\includegraphics{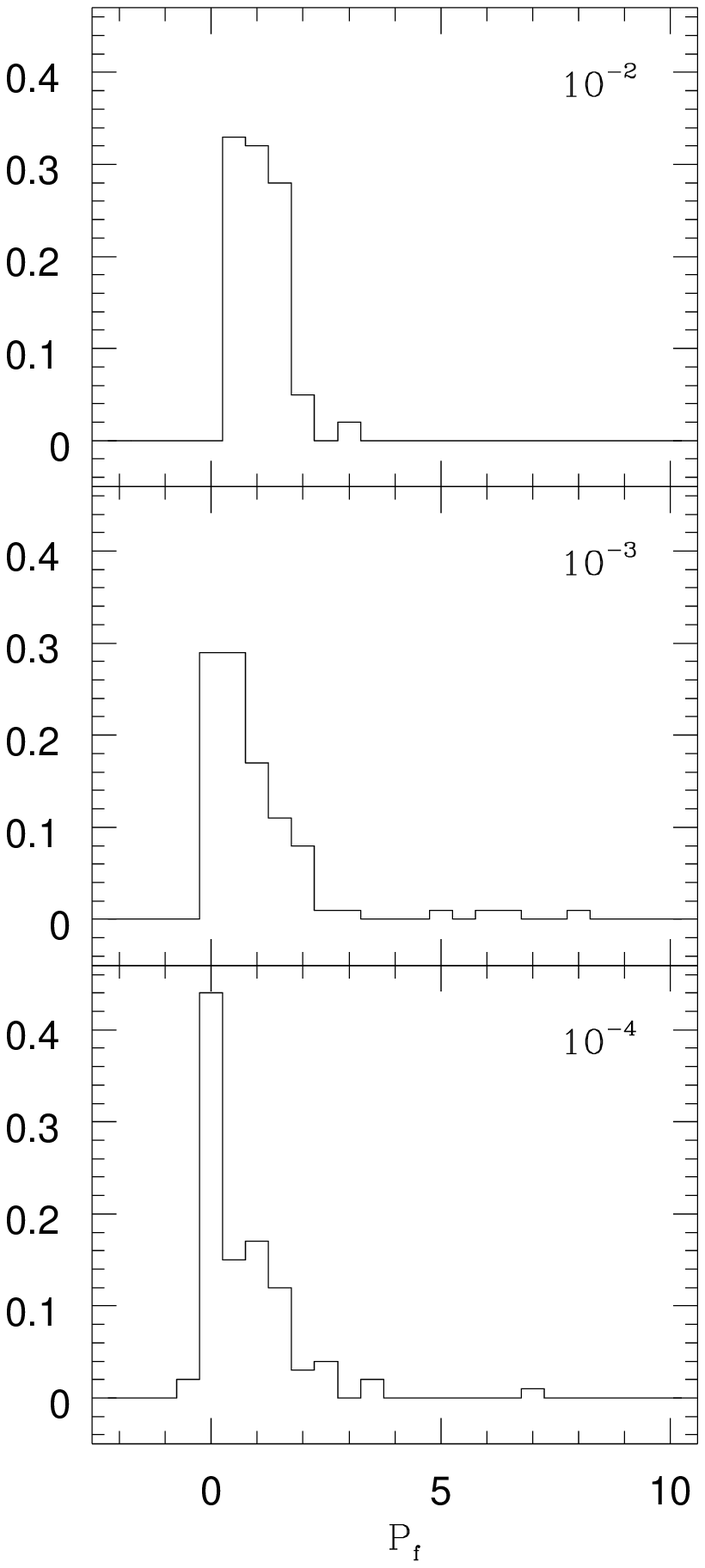}
\caption[junk]{
Momentum probability distributions following the passage of a chaotic resonance.
For the 2 dimensional system with 
$K=0.17$, $\nu=1$, $a=0.2$ shown in Figure 1b) and 1d),
the final momentum distributions
for 100 particles with initial conditions $P=2$ and $\psi \sim \pi$
are shown for the different drift rates 
$-\dot{b} = 10^{-2},10^{-3}$ and $10^{-4}$.
The Lyapunov time for this resonance is $\sim 2 \pi$.
The dispersion of the distribution is nearly constant
among the three integrations.  However the distribution
is more nearly flat for the faster drift rates and
more triangular for the slower drift rates.  
Following passage of a resonance lacking a stochastic zone,
the momentum probability is a delta function
(as shown in Figure 1c).
}
\end{figure*}

\begin{figure*}
\vspace{10.0cm}
\includegraphics{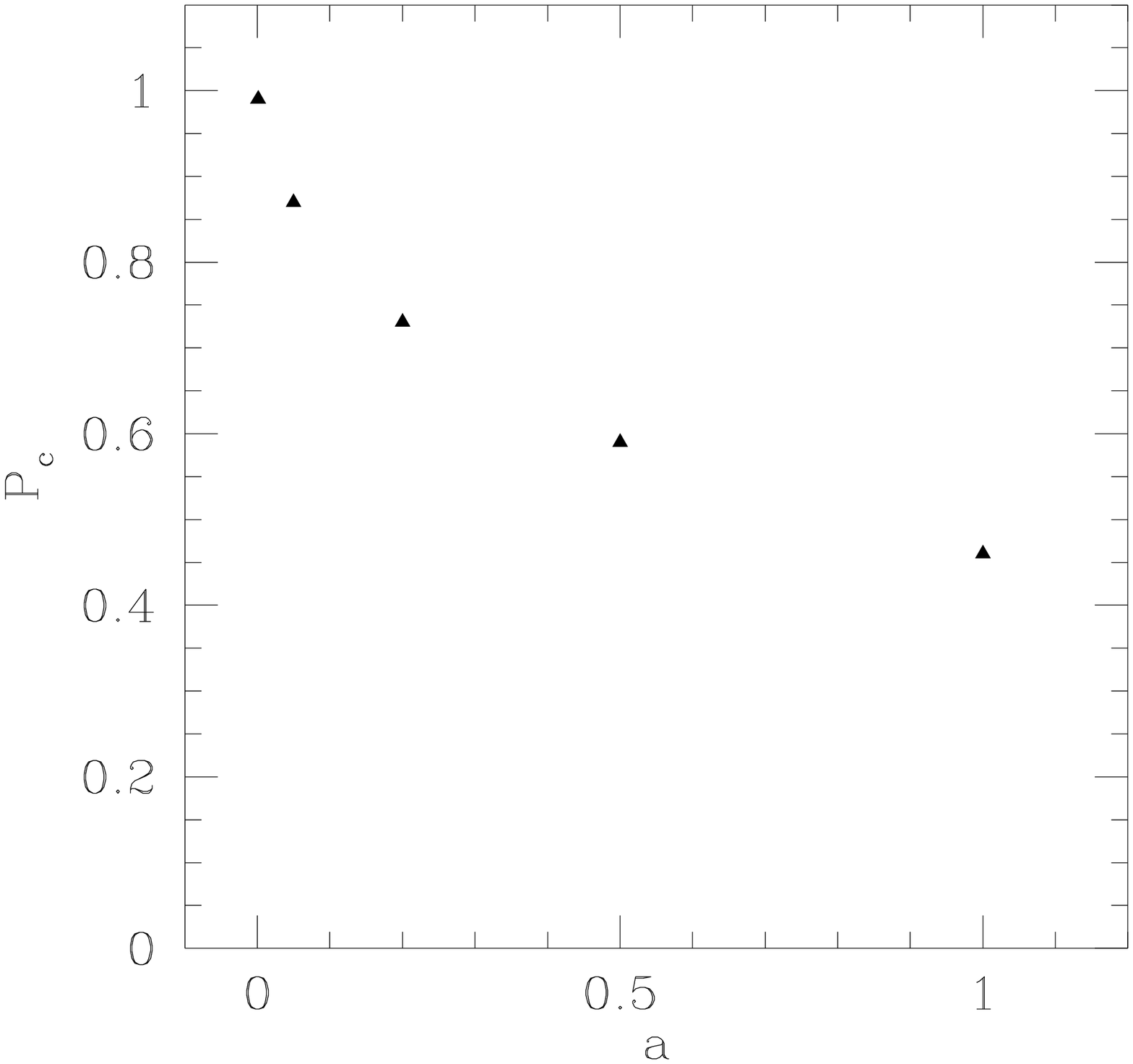}
\caption[junk]{
The capture probability drops as the width of the chaos zone increases.
We plot the capture probability computed 
from simulations each with 100 particles and 
$K=0.17$, $\dot{K}=1.5\times 10^{-3}$, $-\dot{b} = 10^{-3}$, $\nu=1$ 
and $\dot{a}=0$.
The value of $a$ is given as the horizontal axis and the capture probability
on the vertical axis.  The form of the function is expected
if the fraction of the resonance filled with stable islands
drops exponentially as $a \to 0$.
}
\end{figure*}

\begin{figure*}
\vspace{14.0cm}
\caption[junk]{
Numerical integrations of the four dimensional system 
(given by equation 8) are shown for 10 particles.
a) The $I$ momenta as a function of time in units of $10^4$.
b) The $P$ momenta as a function of time in units of $10^4$.
This system has $K=0.17$, $\nu=1$, $a= 0.20$ (similar to
the 2 dimensional system shown in Figure 1, a drift
rate of $-\dot{b} = 10^{-4}$ and $q={1\over 2}$.
Particles were initially set with $P=2$, $I=1$ and randomly distributed
angles.
The  phenomenology of this system is remarkably similar to
that seen in the orbital migration
numerical simulations of Quillen \& Holman (2000).
}
\end{figure*}

\begin{figure*}
\vspace{14.0cm}
\includegraphics{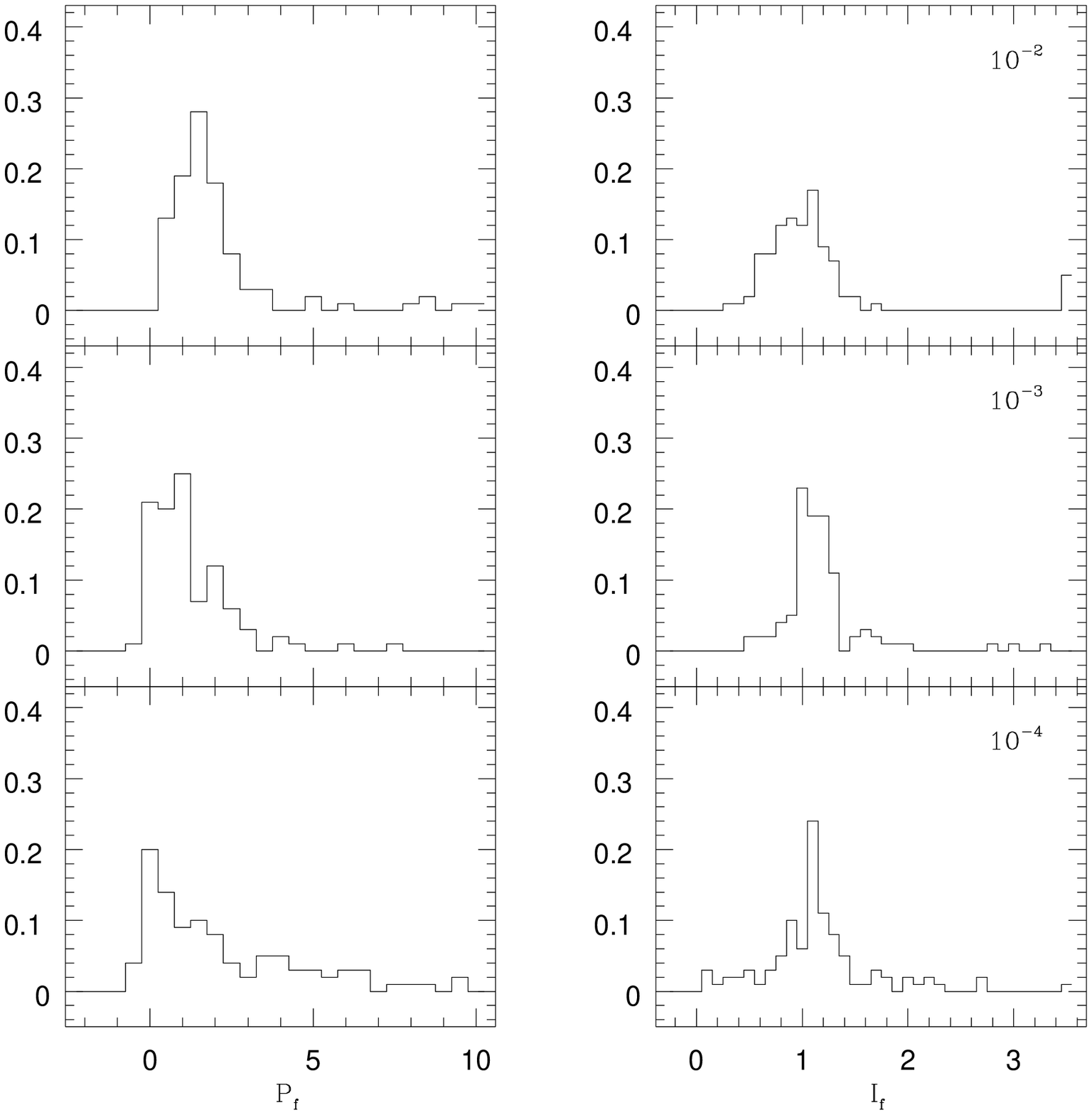}
\caption[junk]{
Momentum probability distributions following passage
of the resonance for the 4 dimensional system described in 
Figure 4 but for three different drift rates  
$-\dot{b} = 10^{-2},10^{-3}$ and $10^{-4}$.
100 particles were integrated for each system.
Drift rates are shown in the upper right of the $I$ plots.
The more slowly drifting system shows larger tails in
both momentum distributions.
}
\end{figure*}

\end{document}